\documentclass[twocolumn,pra,floatfix]{revtex4-2} 
\usepackage{lipsum}
\usepackage[colorlinks,pdfusetitle,urlcolor=blue,citecolor=blue,linkcolor=blue,bookmarksnumbered,plainpages=false]{hyperref}
\usepackage{graphicx}
\usepackage{bm}
\usepackage{amsmath}
\usepackage{amssymb}
\usepackage{braket}
\usepackage{leftidx}
\usepackage{placeins}

\usepackage{scalerel}

\usepackage{subfig}

\renewcommand{\vec}[1]{\bm{#1}}

\newcommand{\ten}[1]{\mathbb{#1}}
\newcommand{\m}{\mathrm{m}}

\newcommand{\G}{\ten{G}}

\newcommand{\tr}{\operatorname{Tr}}

\newcommand{\dif}{\mathrm{d}}
\newcommand{\mi}{\mathrm{i}}

\newcommand{\im}{\mathrm{Im}}

\newcommand{\ra}{ \vec{r}_\mathrm{D}}
\newcommand{\rb}{ \vec{r}_\mathrm{A}}
\newcommand{\D}{\mathrm{D}}
\newcommand{\A}{\mathrm{A}}

\newcommand{\F}{\mathrm{F}}
\newcommand{\lfc}{\mathrm{lfc}}

\newcommand{\re}{\mathrm{Re}}

\newcommand{\e}{ \mathrm{e}}

\newcommand{\Mmm}{M_{\m \m}}
\newcommand{\Mem}{M_{\e \m}}
\newcommand{\Mme}{M_{\m \e}}
\newcommand{\Mee}{M_{\e \e}}
\newcommand{\Gmm}{\G_{\m \m}}
\newcommand{\Gem}{\G_{\e \m}}
\newcommand{\Gme}{\G_{\m \e}}
\newcommand{\Gee}{\G_{\e \e}}

\newcommand{\wA}{\omega_\mathrm{A}}
\newcommand{\wD}{\omega_\mathrm{D}}
\newcommand{\nn}{\nonumber \\}

\newcommand{\nablaL}{\overrightarrow{\nabla}}
\newcommand{\nablaR}{\overleftarrow{\nabla}}

\usepackage{amsfonts} %% <- also included by amssymb
\DeclareMathSymbol{\shortminus}{\mathbin}{AMSa}{"39}
\begin{document}

\title{Macroscopic quantum electrodynamics theory of resonance energy transfer involving chiral molecules}

\author{Janine Franz}
\email{janine.franz@physik.uni-freiburg.de} 
\affiliation{Physikalisches Institut, Albert-Ludwigs-Universit\"at
Freiburg, Hermann-Herder-Str. 3, 79104 Freiburg, Germany}

\author{Stefan Yoshi Buhmann}\email{stefan.buhmann@uni-kassel.de}
\affiliation{Institut für Physik, Universit\"at Kassel, Heinrich-Plett-Straße 40, D-34132 Kassel, Germany}

\author{A. Salam} 
\email{salama@wfu.edu}
\affiliation{Department of Chemistry, Wake Forest University, Winston-Salem, North Carolina 27109, USA}

\date{\today}

\begin{abstract}
Resonance energy transfer between chiral molecules can be used to discriminate between different enantiomers. The transfer rate between chiral molecules consists of a non-discriminatory and discriminatory parts. We derive these two rate contributions in the framework of macroscopic quantum electrodynamics. We show that their ratio is usually larger in the retarded regime or far-zone of large separation distances and that the degree of discrimination can be modified when considering a surrounding medium. We highlight the importance of local field effects onto the degree of discrimination and predict for general identical chiral molecules the optimum dielectric medium for discrimination. We apply our results on to 3-methylcyclopentanone and show that exotic media can even invert the discriminatory effect.
\end{abstract}
%
%\pacs{
%34.35.+a,  % Interactions of atoms and molecules with surfaces
%33.55.+b,  % Optical activity and dichroism
%11.30.Er,  % Charge conjugation, parity, time reversal, and other
%                 % discrete symmetries
%42.50.Nn   % Quantum optical phenomena in absorbing, amplifying,
%                  % dispersive and conducting media; cooperative
%                  % phenomena in quantum optical systems
%}

\maketitle

%%%%%%%%%%%%%%%%%%%%%%%%%%%%%%%%%%%%%%%%%%%%%%%%%%%%%%%%%%%%%%%%%%%%%%
%%%%%%%%%%%%%%%%%%%%%%%%%%%%%%%%%%%%%%%%%%%%%%%%%%%%%%%%%%%%%%%%%%%%%%
\section{Introduction}
A key feature of macroscopic quantum electrodynamics (QED) \citep{Buhmann2007,Scheel2008} that proves
advantageous when it is deployed is that it is able to treat objects that are large relative to the
atomic scale such as plates, slabs and other geometrical bodies. Another benefit is that it also
accounts for the presence of an environment. The surroundings, for example, may be taken to be
purely electric or magnetic, or a combination of the two as in a magneto-dielectric medium, or
even be chiral.
Noteworthy successful early applications of the formalism included the calculation of
Casimir--van der Waals forces \citep{DF1,DF2,Safari2020}. Recently, a number of other problems that require the use of
the quantum properties of light have been tackled. Some of these have covered the generation of
hybrid light-matter (polaritonic) states via ultrastrong coupling, including those due to the presence
of a cavity, and its influence on chemical reactivity \citep{Du2018,Mandal2019,Herrera2020,Basov2021,Li2022,Fregoni2022,Yuen-Zhou2022}, the simulation of molecular emission
power spectra \citep{Wang2019,Wang2020,Feist2021}, modelling Auger decay and interatomic Coulombic decay (ICD) \citep{Hemmerich2018,Bennett2019,Jahnke2020,Cederbaum2021,Franz2021,Franz2022},
as well as predicting discriminatory optical forces occurring between chiral systems \citep{Salam2006,Bradshaw2014,Cameron2014,Forbes2015,Barcellona2017,Suzuki2019,Genet2022}.
Another important inter-particle process that has been considered, and which is of wideranging scientific and technological interest, is resonance energy transfer (RET)
\citep{Avanaki2018,Dung2002}, in what
was an early application in chemical physics of the polariton concept \citep{Knoester1989,Juzeliunas1994}. Subsequent effort
has explored the role of a third body in mediating migration of energy in a dielectric medium, the
different pathways that may ensue, and the interesting coherence/decoherence effects this gives
rise to \citep{Ford2019,Green2020,Waller2022}. This has extended a large body of work (see the references cited in a couple of
recent reviews \citep{Salam2018,Jones2019}) in which the RET phenomenon has been evaluated and fully understood
using molecular QED theory \citep{craigBook,SalamBook,Salam2015,Andrews2018}, where in contrast to macroscopic QED, the Maxwell fields
propagate and are quantized in free space instead of in a medium. Conveyance of electronic energy
takes place via the exchange of a single virtual photon \citep{Andrews2020} between emitter and absorber species,
and the rate is computed perturbatively using the Fermi golden rule. 

A novel feature emerges on
relaxing the common electric dipole approximation. Including the magnetic dipole coupling term enables a discriminatory contribution to the pair transfer rate to occur which applies to migration
of energy between two chiral molecules \citep{Craig1998,Salam2005}. Replacing one of the enantiomers of the pair by
its anti-podal form changes the sign of the discriminatory contribution.

In this work we employ the formalism of macroscopic QED to calculate the RET rate
between a chiral donor and a chiral acceptor in a magnetodielectric medium. On choosing
conditions appropriate to a vacuum electromagnetic field, previous free-space QED results are
recovered. Screening corrections and local field effects are taken into account by utilising a real
cavity model to treat the influence of a surrounding medium, extending electric dipole-dipole
transfer in a dielectric medium \citep{Juzeliunas1994} to the important case of chiral molecules exchanging energy
in a complex environment. Interestingly, it is found that the discriminatory rate may be enhanced
in a magnetodielectric medium. Conditions when this occurs are investigated. The theory
developed is applied to the chiral molecule 3-methylcyclopentanone (3MCP).
After a brief presentation of the macroscopic QED formalism tailored to the RET problem,
the transition matrix element and rate for transfer between two optically active molecules in a
magnetodielectric medium are derived in Sec.~\ref{sec:ret}. The free-space discriminatory rate is obtained and the degree of discrimination quantified in Sec.~\ref{sec:vac}. In the next section, \ref{sec:med}, we investigate the variation of the degree of discrimination as a
function of separation distance and differing medium characteristics to model a number of different
solvents. Conclusions are briefly given in Sec.~\ref{sec:conclusion}.
\section{Resonance energy transfer rate}
\label{sec:ret}
\begin{figure*}[t]
\includegraphics[width = 0.8 \linewidth]{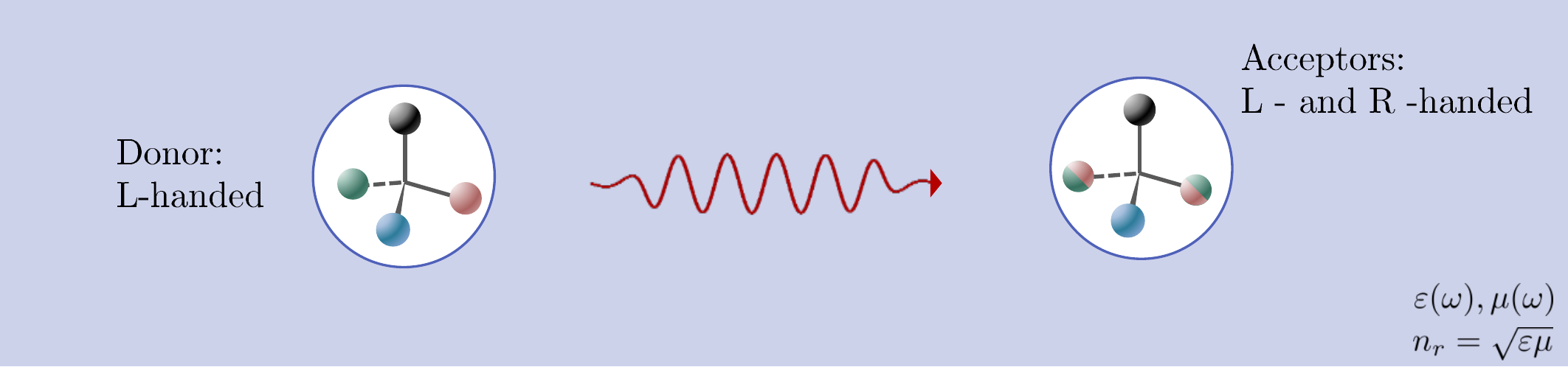} 
\caption{Scheme of the considered setup: Two chiral molecules, one excited (donor) and one in its ground state (acceptor), undergo resonance energy transfer. The donor's handedness is assumed to be known while the acceptor's handedness could be either left or right. We place the system inside a medium and take local field effects and screening corrections around donor and acceptor into account.}
\label{fig:schemes}
\end{figure*}   
We derive, within the framework of macroscopic QED, the resonance energy transfer rate between a chiral donor molecule, $\D$, and a chiral acceptor, $\A$, in a medium (see Fig.~\ref{fig:schemes}). In QED parlance, RET arises from the exchange of a single virtual
photon (or polariton in the case of exchange occurring in a medium) between the two particles \citep{Salam2018}.
The total Hamiltonian operator for the system, in which $\D$ is initially excited in the state $\ket{1}_\D$ and $\A$ is in its ground state $\ket{0}_\A$, is given by 
\begin{align}
\hat{H} 
&= 
\hat{H}_\D + \hat{H}_\A + \hat{H}_\F + \hat{H}_\mathrm{ia}
,
\end{align}
where $\hat{H}_{\D/\A}$ is the particle Hamiltonian for $\D/\A$, $\hat{H}_\F$ is the radiation field Hamiltonian, and the last term denotes the interaction Hamiltonian for the coupling of the electromagnetic field to each molecule.
Because we assume each species to be optically active, the magnetic as well as the usual electric dipole couple to the magnetic field $\hat{\bm B}$ and electric field $\hat{\bm E}$ as
%in the long-wavelength approximation 
\begin{align}
\hat{H}_\mathrm{ia} = - \sum _{\alpha = \D,\A} \left[ \hat{\bm d}^{(\alpha)} \cdot \hat{\bm E}(\bm r_\alpha) + \hat{\bm m}^{(\alpha)} \cdot \hat{\bm B}(\bm r_\alpha) \right]
,
\label{Hint}
\end{align}
where $\bm r_{\D/\A}$ is the donor's/acceptor's position, $\hat{\bm d}^{\D/\A}$ is the donor's/acceptor's electric dipole moment operator and $\hat{\bm m}^{\D/\A}$ is its magnetic counterpart.
The energy migration rate $\Gamma$ is calculated using Fermi's golden rule,
\begin{align}
\Gamma &=
\frac{2 \pi}{\hbar^2} \rho(\omega_f) 
\big|  M_{fi}\big|^2 
,
\label{gen:FGR}
\end{align}
where $M_{fi}$ is the matrix element for the transition between initial and final state, and $\rho(\omega_f)$ is the density of final states with energy $E_f = \hbar \omega_f$. To leading order, $M_{fi}$ is evaluated from the second-order perturbation theory formula
\begin{align}
M_{fi} &= 
\sum_{j} \frac{\braket{f | \hat{H}_{I} | j} \braket{j| \hat{H}_{I} | i}}{E_i - E_j}
\Big|_{E_i = E_f}
,
\label{Mfi}
\end{align}
where $E_{x}$ is the energy of the respective state $\ket{x}$ and for the system of interest the initial and final states are given by: $\ket{i} = \ket{1}_\mathrm{D} \ket{0}_\mathrm{A} \ket{ \{ \bm 0 \} }_F$ and $\ket{f} = \ket{0}_\mathrm{D} \ket{1}_\mathrm{A} \ket{ \{ \bm 0 \} }_F $. Our analysis focusses on only one donor and acceptor pair. The extension to $N$ particles is however straightforward. By summing over all possible final states one can account for $n$ ground state acceptors and $m$ excited donors. Only if multiple donors share an excitation does the calculation become more involved and superradiance effects occur \citep{Bang2019}.

From Eq.~\eqref{Mfi}, it is necessary to sum over all possible intermediate states 
\begin{align}
\ket{j} 
&\in \Big\{ 
 \ket{j_1 (\lambda, \omega,\bm r) } ,
 \ket{j_2(\lambda, \omega,\bm r) }
 ;\, \forall \lambda, \bm r, \omega 
\Big\} 
,
\end{align}
with 
\begin{align}
\ket{j_1 (\lambda, \omega,\bm r) } &= \ket{0}_\mathrm{D} \ket{0}_\mathrm{A} \ket{  \bm 1_\lambda(\omega,\bm r) }_F 
,
\\
  \ket{j_2(\lambda, \omega,\bm r) } &= \ket{1}_\mathrm{D} \ket{1}_\mathrm{A} \ket{ \bm 1_\lambda(\omega,\bm r) }_F
, 
\end{align}
where $\ket{0/1}_{\D/\A}$ is the ground/excited state of the donor/acceptor molecule and $\ket{\bm 1_\lambda (\omega, \bm r)}_F = \hat{\bm f}^\dagger_\lambda( \bm r, \omega) \ket{\{ 0 \}}_F$ is the single-quantum Fock state of collective, polariton-like bosonic excitations of electric/magnetic type ($\lambda = \e/\m$) at position $\bm r$ with energy $\hbar \omega$. The creation and annhilation operators $\hat{\bm f}^{\dagger}_\lambda(\bm r, \omega)$ and $\hat{\bm f}_\lambda(\bm r, \omega)$ fulfil the commutation relations  
\begin{align}
\left[ 
\hat{\bm f}_\lambda (\bm r , \omega) ,\hat{\bm f}^\dagger_{\lambda'} (\bm r', \omega') 
\right]
=
\delta_{\lambda \lambda'} \bm \delta(\bm r - \bm r') \delta( \omega  - \omega')
,
\end{align}
and may be used to expand the electric and magnetic fields as 
\begin{align} 
 \hat{\bm E} (\bm r)
 &= 
 \int_0 ^\infty \dif \omega \hat{\bm E}(\bm r, \omega) + \mathrm{h.c.}
,
\label{Efield}
\\
 \hat{\bm E} (\bm r, \omega) 
 &= 
 \sum_{\lambda=\e,\m} \int \dif^3 r'  \G_\lambda ( \bm r, \bm r', \omega ) \cdot \hat{\bm f}_\lambda  (\bm r', \omega)  
,
\label{Ewfield}
 \\
 \hat{\bm B} (\bm r)
 &= 
 \int_0 ^\infty \dif \omega \hat{\bm B}(\bm r, \omega) + \mathrm{h.c.}
,
 \\
 \hat{\bm B} (\bm r, \omega) 
 &= \frac{1}{\mi \omega}
 \sum_{\lambda=\e,\m} \int \dif^3 r' \nablaL \times \G_\lambda ( \bm r, \bm r', \omega ) \cdot \hat{\bm f}_\lambda  (\bm r', \omega)  
.
\label{Bwfield}
\end{align}
\begin{figure}[b]
\includegraphics[width = 0.85 \linewidth]{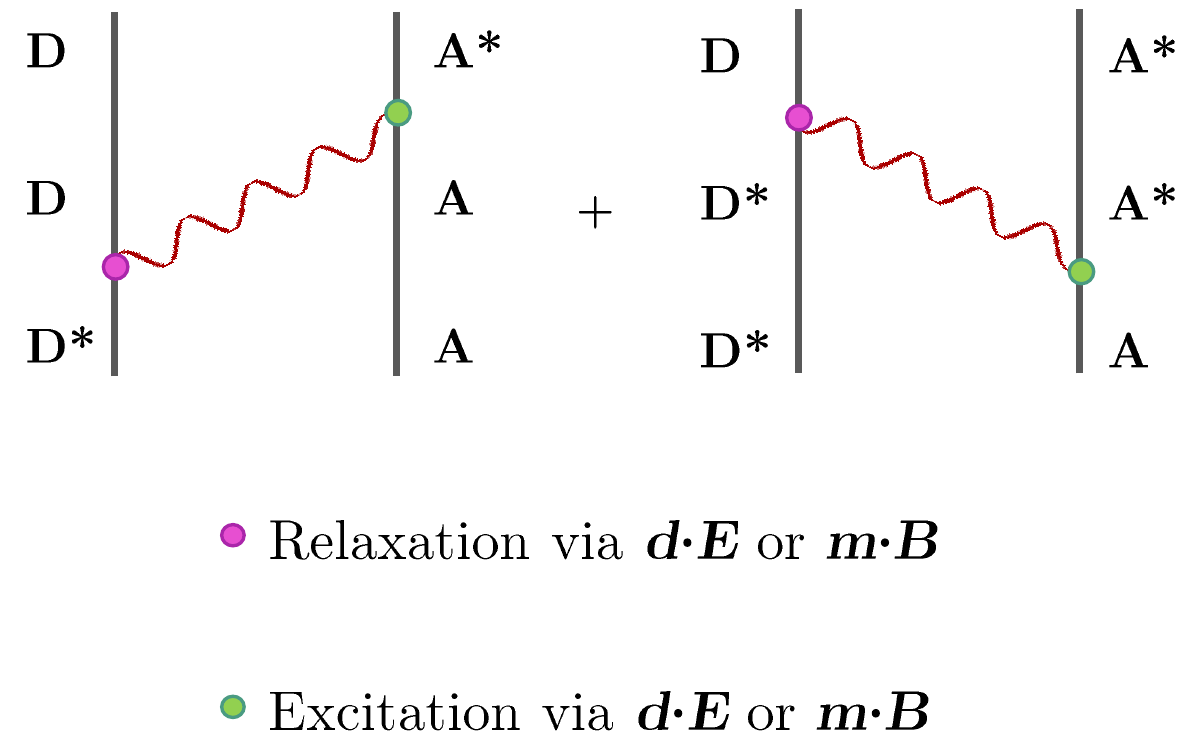} 
\caption{Feynman diagrams for RET between chiral particles: Each intersection between a wavy and solid line represents a particle-field-interaction. Each interaction can either be electric or magnetic. The two diagrams result in different frequency poles (see Eq.~\eqref{Mem-calc}).}
\label{fig:feynman}
\end{figure} 
The Green's tensor $\G_\lambda$ is defined via the Helmholtz equation and is given explicitly in Appendix \ref{app:G}.
The RET process may be visualized by the two Feynman diagrams of Fig.~\ref{fig:feynman}. They reflect the two possible propagation directions associated with single virtual photon exchange between emitter and absorber. Each intersection of a solid line with a wavy line represents an interaction between the respective particle and the field. Based on the coupling Hamiltonian, Eq.~\eqref{Hint}, electric $\left( - \hat{\bm d }\cdot \hat{\bm E} \right)$ as well as magnetic $\left( - \hat{\bm m }\cdot \hat{\bm B} \right)$ dipole interactions with the Maxwell field operators have to be considered. The probability amplitude can therefore be divided into four terms 
\begin{align}
M_{fi}
&= 
 \Mee + \Mem + \Mme +\Mmm
,
\end{align}
where the first and second subscript imply electric or magnetic interaction of $\A$ and $\D$ with the field, respectively. 
By using the vacuum correlation functions of the electromagnetic fields (see Appendix \ref{app:sec:Mfi}),
\begin{multline}
\braket{\hat{\vec{E}}  (\bm r_\D, \omega ) \otimes \hat{\vec{E}}^\dagger  (\bm r_\A, \omega' ) }_\mathrm{vac}
 =
 \\ 
 \frac{\hbar \mu_0 \omega^2 }{\pi}  \delta(\omega -\omega') \im \G (r_\A,r_\D,\omega) 
 \label{EEcorr}
\end{multline}

\begin{multline}
\braket{\hat{\vec{E}} (\bm r_\D, \omega ) \otimes \hat{\vec{B}}^\dagger (\bm r_\A, \omega' ) }_\mathrm{vac}
= 
\\
- \frac{\mi \hbar \mu_0  \omega }{\pi }    \delta(\omega -\omega')   \im \G (r_\A,r_\D,\omega)  \times \nablaR_\D   
 \label{EBcorr}
\end{multline} 
\begin{multline}  
\braket{ \hat{\vec{B}} (\bm r_\D, \omega ) \otimes \hat{\vec{E}}^\dagger (\bm r_\A, \omega' )  }_\mathrm{vac}  =
\\
= 
- \mi  \frac{\hbar \mu_0 \omega }{\pi} \delta( \omega - \omega') \nablaL_\A \times \im \G (r_\A,r_\D,\omega)
 \label{BEcorr}
\end{multline} 
\begin{multline}
\braket{\hat{\vec{B}} (\bm r_\D, \omega ) \otimes \hat{\vec{B}}^\dagger (\bm r_\A, \omega' ) }_\mathrm{vac}
= 
\\
- \frac{\hbar \mu_0  }{\pi }    \delta(\omega -\omega') \nablaL_\A \times \im \G (r_\A,r_\D,\omega)  \times \nablaR_\D   
,
\label{BBcorr}
\end{multline}
we can derive the transition matrix elements in terms of the Green's tensor for general environments as (see Appendix \ref{app:sec:Mfi})
\begin{align}
\Mem &= 
- \frac{1}{\hbar } \! \int \!\! \dif \omega 
\Big\lbrace  
\frac{1}{\omega -  \wD} \vec{d}^\A \cdot \braket{  \vec{\hat{E}}_\A \otimes \vec{\hat{B}}'^{\dag}_\D   }_\mathrm{vac} \cdot \vec{m}^\D
\nn
& \qquad\qquad\quad
+ 
\frac{1}{\omega +  \wA} \vec{m}^{\D}\cdot \braket{  \vec{\hat{B}}_\D \otimes \vec{\hat{E}}'^{\dag}_\A }_\mathrm{vac} \cdot \vec{d}^\A
\Big\rbrace
\nn
&= 
\frac{\mu_0 c^2}{\pi}
\! \int \!\! \dif \omega  
 \big\{ 
 \frac{1}{\omega -  \wD} + \frac{1}{\omega + \wD}
 \big\}
\nn
& \qquad 
\bm d^\A \cdot
\im \G (r_\A,r_\D,\omega)  \times \nablaR_\D \cdot \bm m ^\D
,
\label{Mem-calc}
\end{align}
where $\wD = \omega^\mathrm{D}_1 - \omega^\mathrm{D}_0$  is the transition frequency of the donor, and the acceptor's transition frequency is $\wA =  \omega^\mathrm{A}_1 - \omega^\mathrm{A}_0 = \wD$ from energy conservation, 
and we introduced the shorthand-notation $\hat{\bm E} _{\A/\D} = \hat{\bm E} (\bm r_{\A/\D} , \omega  )$, analogously for $\hat{ \bm B}$, while the prime indicates the substitution of $\omega \rightarrow \omega'$ and the arrows on the the Nabla operator denote the direction which the derivative acts on. 
Additionally, we introduced the following notation  for the downward and upward dipolar transition in $\D$ and $\A$, respectively: 
$\bm d^\mathrm{D} = \braket{0 | \hat{\bm d} | 1}_\mathrm{D}$ and $\bm d^\mathrm{A}= \braket{1| \hat{\bm d} | 0}_\mathrm{A}$, with analogous definitions for $\bm m^{\D/\A}$.
The remaining transition matrix elements can be derived similarly and the pole-integration can be calculated generally as
\begin{multline}
\int 
 \!\! \dif \omega 
\Big\lbrace  
\frac{
f(\omega) }{\omega -  \wD}  
+ 
\frac{
f(- \omega) }{\omega +  \wA}
\Big\rbrace \im \G(\omega)
%\\
%\times
%\frac{1}{2 \mi }
%\left( \G(\omega ) + \G(- \omega ) \right)
\\
\longrightarrow
\lim_{\epsilon \rightarrow 0^+}
\int 
 \!\! \dif \omega 
\Big\lbrace  
\frac{
f(\omega) }{\omega -  ( \wD + \mi \epsilon) }  
+ 
\frac{
f(- \omega) }{\omega +  \wA}
\Big\rbrace
\\
\phantom{hjdklashdljdslaadlas}\times
\frac{1}{2 \mi }
\left( \G(\omega ) + \G(- \omega ) \right)
\\
= 
\pi f(\wD ) \G (\wD )
,
\label{poleInt}
\end{multline}
for $\wA = \wD$ and $f(\omega) = \omega^n$ with $n \in \{0,1,2\}$.  The correct regularisation of the pole on the real axis follows from revisiting the derivation of Fermi's golden rule and the detailed pole integration is given in the Appendix~\ref{app:pole}
This finally yields the desired transition matrix elements 
\begin{subequations}
\begin{align}
\Mee
&=
- 
\mu_0 \wD^2
\bm d^\mathrm{A}\cdot    \G(\bm r_\mathrm{A}, \bm r_\mathrm{D}, \wD)  \cdot \bm d^\mathrm{D}
,
\label{gen:Mee-nondual}
\\
\Mem
&=
\mi \mu_0  \wD
\bm d^\mathrm{A}\cdot\G(\bm r_\mathrm{A}, \bm r_\mathrm{D}, \wD)  \times \nablaR_\D \cdot \bm m^\mathrm{D}
,
\\
\Mme
&=
\mi \mu_0 \wD
\bm m^\mathrm{A}\cdot \nablaL_\A \times  \G(\bm r_\mathrm{A}, \bm r_\mathrm{D}, \wD) \cdot \bm d^\mathrm{D}
,
\\
\Mmm
&=
\mu_0 
\bm m^\mathrm{A}\cdot \nablaL_\A\times \G(\bm r_\mathrm{A}, \bm r_\mathrm{D}, \wD) \times \nablaR_\D \cdot \bm m^\mathrm{D}
.
\label{gen:Mmm-nondual}
\end{align}
\end{subequations} 
One can easily verify that they lead to the same results known from free-space QED \citep{Craig1998}. 
In vacuum, the Green's tensor is given by (see Appendix \ref{app:G}) 
\begin{align}
\G^{(0)}(\bm r_\mathrm{A},\bm r_\mathrm{D},\omega) = \left[\ten{I} + \frac{ \nabla \nabla }{k^2} \right] \frac{e^{\mi k r}}{4 \pi r}
,
\end{align}
with $r = |\bm r_\mathrm{D} - \bm r_\mathrm{A}|$, $k = \omega/c$, $\ten{I}$ is the 3x3-identity matrix, $\nabla = \nablaL$ and $\nabla \nabla = \nabla \otimes \nabla $ . For dipoles of the same type at each particle, familiar matrix elements ensue~\citep{SalamBook},
\begin{align}
\Mee
&= -\mu_0 c^2 d^\mathrm{A}_i \left[ k^2 \delta_{ij} + \nabla_i \nabla_j \right] \frac{e^{\mi k r}}{4 \pi r} d^\mathrm{D}_j 
\nn
&=   \frac{1}{4\pi \varepsilon_0} d^\mathrm{A}_i  d^\mathrm{D}_j  \left[ \nabla^2 \delta_{ij}- \nabla_i \nabla_j \right] \frac{e^{\mi k r}}{ r}
,
\label{gen:Mee-consistency}
\end{align}
\begin{align}
\Mmm
&= 
\mu_0 m_i^\mathrm{A} \Big[ 
- \nabla  \times  \ten{I} \times \nabla  
\nn& \qquad\qquad\qquad 
+ \nabla  \times \frac{\nabla  \nabla  }{k^2} \times \nabla     \Big]_{ij} 
\frac{e^{\mi k r}}{4 \pi r} m_j^\mathrm{D}
\nn
&= 
- \mu_0 m_i^\mathrm{A}  m_j^\mathrm{D} \left[  \nabla  \times  \ten{I} \times \nabla     \right]_{ij} \frac{e^{\mi k r}}{4 \pi r}
\nn
&= 
\frac{1}{4\pi \varepsilon_0 c^2 } m_i^\mathrm{A}  m_j^\mathrm{D} \left[  \nabla^2  \delta_{ij}  - \nabla_i \nabla_j \right]  \frac{e^{\mi k r}}{ r}
,
\label{gen:Mmm-consistency}
\end{align}
while for the mixed dipole term, crucial for transfer between chiral systems \citep{Craig1998}
\begin{align}
\Mem + \Mme
&=  
\frac{\mi k}{ 4\pi \varepsilon_0 c}
\left( 
d^\mathrm{D}_i  m^\mathrm{A}_j  -  m^\mathrm{D}_i d^\mathrm{A}_j 
\right)
\epsilon_{ijk} \nabla_k \frac{e^{\mi k r}}{ r}
,
\label{gen:Mem-consistency}
\end{align}
where we used
\begin{align}
\begin{split}
\left[ \nablaL_\A\times \G(r_\mathrm{A},r_\mathrm{D},\omega) \right]_{ij}
&= 
- \epsilon_{ijk} \nabla_k \frac{e^{\mi k r}}{4 \pi r}
\\ 
&= - 
\left[ \G(r_\mathrm{A},r_\mathrm{D},\omega) \times \nablaR_\A  \right]_{ij}
\end{split} .
\end{align}
Inserting the matrix elements in terms of the general Green's tensor \eqref{gen:Mee-nondual}--\eqref{gen:Mmm-nondual} into Fermi's golden rule \eqref{gen:FGR}, leads to the rate for a general environment
\begin{align} 
\Gamma &=  \sum_{\lambda_1,\lambda_2,\lambda_3,\lambda_4} \Gamma_{\lambda_1\lambda_2\lambda_3\lambda_4}
,
\\
\Gamma_{\lambda_1\lambda_2\lambda_3\lambda_4}
&=
\frac{2 \pi \rho \mu_0^2 }{9 \hbar^2} ( \bm d^\mathrm{A}_{\lambda_1}\cdot \bm d^\mathrm{A*}_{\lambda_2} ) 
 ( \bm d^\mathrm{D*}_{\lambda_3}\cdot \bm d^\mathrm{D}_{\lambda_4} ) 
 \nn
 & \qquad\qquad\qquad\qquad \times
 \tr \left[ \G_{\lambda_1 \lambda_4}\cdot \G^{*T}_{\lambda_2 \lambda_3}\right]
,
 \label{dual:rate} 
\end{align}
where we have assumed that the transitions are isotropic, such that $\bm d_1 \otimes \bm d_2 = (\bm d_1 \cdot \bm d_2) \ten{I} /3$ and we adopted a dual formulation with  $\lambda_i \in \{ \e , \m \}$. The relevant dual quantities are
\begin{subequations}
\begin{align}
\bm d_\mathrm{e} &= \bm d, \qquad \bm d_\mathrm{m} = \frac{\bm m }{c}
,
\label{dualdef1}
\\
\Gee &= \frac{\mi \omega}{c} \G (\rb, \ra , \omega ) \frac{\mi \omega}{c}
,
\label{Gee}
\\
\Gmm &= \nablaL_\A \times \G (\rb,\ra, \omega ) \times \nablaR_\D
,
\\
\Gem &=\frac{\mi \omega}{c}\G (\rb,\ra, \omega ) \times \nablaR_\D
,
\\
\Gme &= \nablaL_\A \times \G (\rb,\ra, \omega )  \frac{\mi \omega}{c}
,
\label{Gme}
\end{align}
\end{subequations}
where $\hbar \omega = \hbar \omega^\D = \hbar \omega^\A$ is the transition energy. 
When the handedness of one participating molecule is known, the rate can be used to discriminate between different enantiomers of the second entity. Without loss of generality, 
we take the donor to be left-handed while the acceptor species may be of either handedness -- see Fig.~\ref{fig:schemes}. 
Molecular chirality may be characterised by the scalar product  between magnetic and electric transition dipole moments, which
is related to the rotatory strength $R$ of the respective chiral molecule through,
\begin{align}
\frac{R}{c}  &= \im\left[ \braket{0| \hat{ \bm d}_\mathrm{e} | 1} \cdot \braket{1| \hat{ \bm d}_\mathrm{m}| 0}    \right],
\end{align}
and whose sign depends on the molecule's handedness.
In our notational convention, where $\bm d^\mathrm{D}_\lambda$ denotes a downward transition, while $\bm d^\mathrm{A}_\lambda$ represents an upward transition, the respective rotatory strengths of $\D$ and $\A$ are given by
\begin{align}
\frac{R^\mathrm{D}}{c} 
&= \im \left[ \bm d^\D_\mathrm{e} \cdot \bm d^{\D*}_\mathrm{m} \right] 
= - \mi \bm d^\D_\mathrm{e} \cdot \bm d^{\D*}_\mathrm{m} 
= \mi \bm d^{\D*}_\mathrm{e} \cdot \bm d^\D_\mathrm{m} 
,
\\
\frac{R^\mathrm{A}}{c} 
&= \im \left[ \bm d^{\A*}_\mathrm{e} \cdot \bm d^\A_\mathrm{m} \right] 
= \mi \bm d^{\A}_\mathrm{e} \cdot \bm d^{\A*}_\mathrm{m} 
= - \mi \bm d^{\A*}_\mathrm{e} \cdot \bm d^{\A}_\mathrm{m} 
,
\end{align}
where we have explicitly assumed real electric transition dipole moments $\bm d_\mathrm{e} = \bm d_\mathrm{e}^*$ and imaginary magnetic dipole moments $\bm d_\mathrm{m} =- \bm d_\mathrm{m}^*$.
%Although we only consider one donor and one acceptor molecule here, the extension to larger systems is straight forward: Multiple acceptors can be included by simply summing the rate equation Eq.~\eqref{rate_fgr} over all available final states, multiple donors that not entangled can be similarly take into account by using the respective density matrix as initial state, which leads to a sum of Eq.~\eqref{rate_fgr} over all initial states times their respective probability. Donors that coherently share an excitation can lead however to superradiance.
\section{Discrimination in free space}
\label{sec:vac}
\begin{figure} 
\centering
\includegraphics[width =  \linewidth]{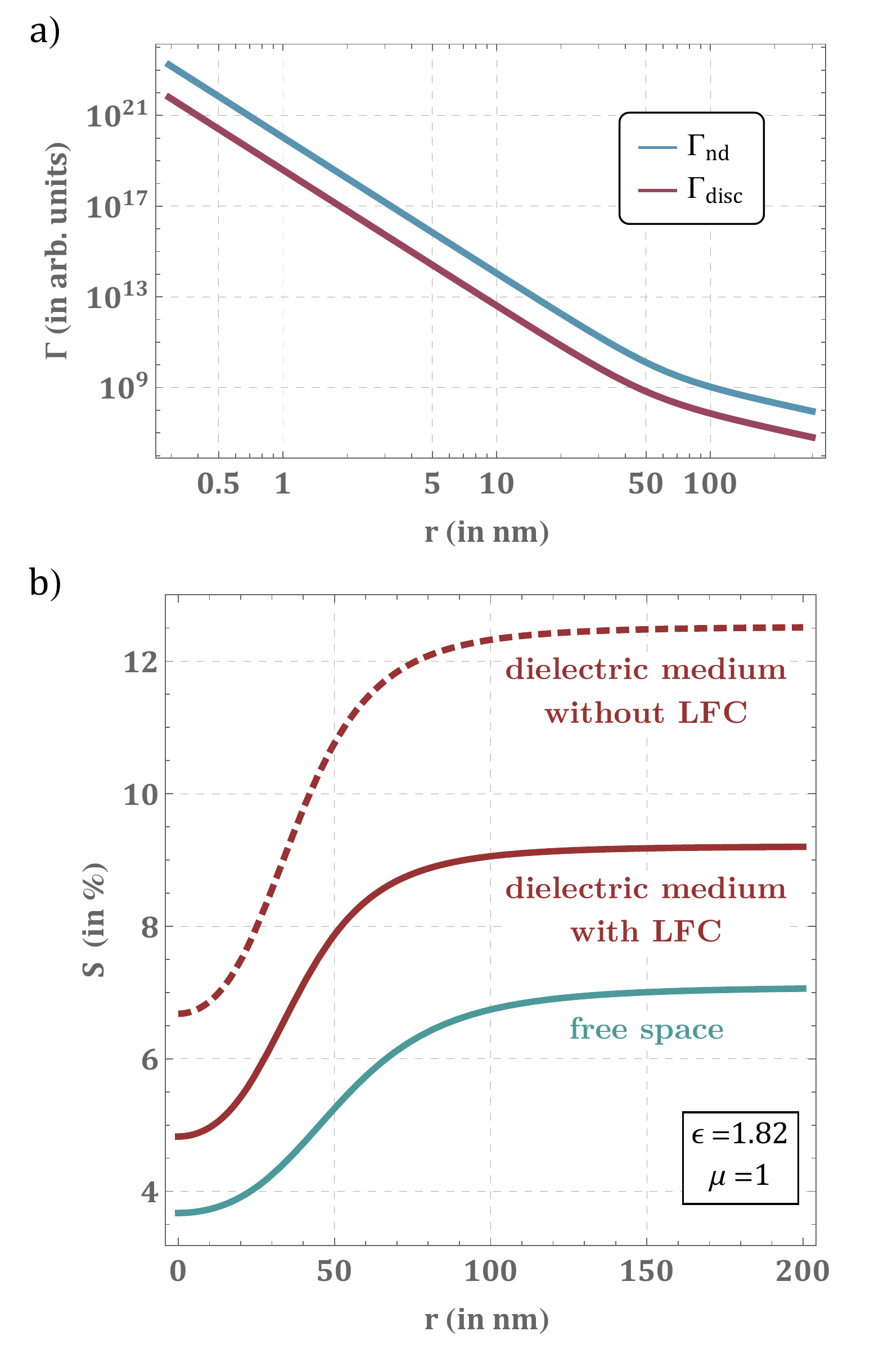}
\caption{a) Discriminatory and non-discriminatory rate contribution for RET in free space between chiral donor and acceptor as a function of their separation distance $r$ on the example of 3MCP. b) Degree of discrimination $S =  \Gamma_\mathrm{disc}/\Gamma_\mathrm{nd}$ for 3MCP as a function of separation distance $r$ in free space, in water without the local field correction and in water with local field correction. 
The full rate contributions \eqref{fs:gammadisc} and \eqref{fs:gammand} are used for the plots.}
\label{fig:rates_and_Svac}
\label{fig:SmedLFC}
\end{figure}
With these preparations at hand, we can partition the rate into different contributions, depending on their sensitivity to the acceptor's handedness. The potentially discriminatory contributions that are proportional to the acceptor's optical rotatory strength $R^\A$ are
\begin{align}
\Gamma_\mathrm{disc} = \sum_{\lambda_1 \lambda_2} \left( \Gamma_{\e \m \lambda_1 \lambda_2} + \Gamma_{\m \e \lambda_1 \lambda_2} \right)
.
\end{align}
These rate contributions are of two different forms
\begin{align}
\Gamma_{\e \m \lambda \lambda} 
&\propto
  R^\mathrm{A}
| \bm d^\D_{\lambda }|^2  \tr \left[ \G_{\e \lambda }\cdot \G^{*T}_{\m \lambda }\right]
,
\label{vanishing}
\\
\Gamma_{\e \m \lambda_1 \lambda_2} 
&\propto
  R^\mathrm{A}
 R^\mathrm{D}  \tr \left[ \G_{\e \lambda_1 }\cdot \G^{*T}_{\m \lambda_2 }\right] 
,
\end{align}
with $\lambda_1 \neq \lambda_2$ and analogously for $\Gamma_{\m \e\lambda \lambda'}$. In agreement with Curie's dissymmetry principle, contributions of the first kind (Eq.~\eqref{vanishing}) vanish in free space: $\tr \left[ \G_{e \lambda  }\cdot \G^{*T}_{m \lambda }\right]  = \tr \left[ \G_{m \lambda  }\cdot \G^{*T}_{e \lambda }\right]= 0 $, $\forall \lambda$. Only contributions that emerge from the chiral properties of both donor and acceptor, i.e.~that are proportional to the product $R_\mathrm{A} R_\mathrm{D}$, can discriminate the acceptor's enantiomers. We are hence left with the discriminatory rate contribution in free space 
\begin{align}
\Gamma_\mathrm{disc}
&= 
\Gamma_{\e \m \m \e} + \Gamma_{\m \e \m \e} + \Gamma_{\e \m \e \m} + \Gamma_{\m \e \e \m}
,
\end{align}
and the non-discriminatory rate contribution 
\begin{align}
\Gamma_\mathrm{nd}
&= \sum_{\lambda_1 \lambda_2} \Gamma_{\lambda_1 \lambda_1 \lambda_2 \lambda_2}
,
\end{align}
which yield the total rate for left($\mathrm{L}$)- and right($\mathrm{R}$)-handed acceptors, $ \Gamma_\mathrm{L/R} = \Gamma_\mathrm{nd} \pm | \Gamma_\mathrm{disc}|$. 
By using the free space Green's tensor \eqref{app:Gfs} we find for the two rate contributions 
\begin{align}
\Gamma_{\text{nd}} 
&= 
\frac{\rho }{36 \pi   \epsilon_0^2 \hbar^2 r^6}
\Bigg\lbrace 
\left(  | \bm d^\mathrm{A}_\mathrm{e}|^2 | \bm d^\mathrm{D}_\mathrm{e}|^2  +   | \bm d^\mathrm{A}_\mathrm{m}|^2 | \bm d^\mathrm{D}_\mathrm{m}|^2  \right) 
\nn
& \qquad\qquad\qquad\qquad \times
\bigg[
3 + 				
 \frac{ \omega^2  r^2 }{c^2 } 
 +
 \frac{ \omega^4  r^4}{c^4 }  
\bigg]  
\nn
&\qquad+
\left(  | \bm d^\mathrm{A}_\mathrm{e}|^2  
 | \bm d^\mathrm{D}_\mathrm{m}|^2 
+   | \bm d^\mathrm{A}_\mathrm{m}|^2  
| \bm d^\mathrm{D}_\mathrm{e}|^2 \right)
\nn
& \qquad\qquad\qquad\qquad\qquad \times
\bigg[
 \frac{ \omega^2  r^2 }{c^2 } 
 +
 \frac{ \omega^4  r^4}{c^4 }  
\bigg]  
\Bigg\rbrace
,
\label{fs:gammand}
\end{align}
\begin{align}
\Gamma_{\text{disc}} 
&=  \rho
\frac{R_\mathrm{D} R_\mathrm{A}}{18 \pi  c^2 \epsilon_0^2 \hbar^2 r^6}
\left(3 +2  \frac{r^2 \omega^2}{c^2}  + 2  \frac{ r^4 \omega^4}{c^4}\right) 
.
\label{fs:gammadisc}
\end{align}
We define the degree of discrimination as 
\begin{align}
S &=
\frac{\Gamma_\mathrm{L} - \Gamma_\mathrm{R}}{\Gamma_\mathrm{L} + \Gamma_\mathrm{R}} = \frac{\Gamma_\mathrm{disc}}{\Gamma_\mathrm{nd}} \in [-1, 1]
.
\end{align}
Usually the magnetic dipole is much smaller than the electric one, i.e.~$\bm d_\mathrm{m} \ll \bm d_\mathrm{e}$. With this approximation the degree of discrimination in free space is given by the simple expression 
\begin{align}
S \approx
\frac{4 R_\mathrm{D} R_\mathrm{A}}{ c^2 | \bm d^\mathrm{A}_\mathrm{e}|^2 | \bm d^\mathrm{D}_\mathrm{e}|^2  }  
\frac{ 3 +2  k_0^2 r^2 + 2  k_0^4 r^4  }{ 3 +  k_0^2 r^2 +   k_0^4 r^4 } 
.
\label{Sfs}
\end{align}
It exhibits a lower bound $S_{r \rightarrow 0}$ in the nonretarded or near-zone limit of small separations and an upper bound $S_{r \rightarrow \infty}$ in the retarded or far-zone limit of large distances. These limits may be derived analytically, yielding
\begin{align}
S_{r \rightarrow 0 } 
&\approx \frac{4 R_\mathrm{D} R_\mathrm{A}}{   c^2| \bm d^\mathrm{A}_\mathrm{e}|^2 | \bm d^\mathrm{D}_\mathrm{e}|^2  }
,
\label{Svacnr}
\\
S_{r \rightarrow \infty} 
&\approx  \frac{8 R_\mathrm{D} R_\mathrm{A}}{   c^2| \bm d^\mathrm{A}_\mathrm{e}|^2 | \bm d^\mathrm{D}_\mathrm{e}|^2  } = 2
S_{r \rightarrow 0 }
.
\label{Svacr}
\end{align}
Although the absolute rate rapidly decreases with increasing separation distance ($\sim r^{-6}$), the discrimination is stronger by a factor of approximately $2$ in the far-zone ($ r \omega/c > 1$). This can bee seen in Fig~\ref{fig:rates_and_Svac}, where the rate contributions and degree of discrimination in free space as well as in a medium are plotted as a function of separation distance $r$ between donor and acceptor for the example of 3-methyl-cyclopentanone (3MCP) as chiral donor and acceptor. 

The transition frequency for 3MCP is given by $\omega = 6.44 \times 10^{15} \mathrm{s}^{-1}$ and as we can see in Fig.~\ref{fig:rates_and_Svac} the retarded limit is reached at separation distances of roughly $r = 2 c/\omega \approx 100$~nm. 

The chosen example of 3MCP features a transition with a very small electric transition dipole ($|\bm d_e| = 2.44\times 10^{-31}$ Cm) compared to its magnetic transition dipole ($|\bm d_m|= 3.31\times 10^{-32}$ Cm)\citep{Kroener2011, Horsch2011,Kroener2015, Suzuki2019}. This leads to a relatively large rotatory strength $R/c = \im[ \bm d_e \cdot \bm d_m]$. 
If we define an angle such that:  $R/c = |\bm d_e| |\bm d_m |\cos\theta $, we find $\cos \theta = 0.98 \approx 1$ for 3MCP. Nonetheless, the maximum degree of discrimination in free space for 3MCP is only at $S = 7$\%.

In Fig.~\ref{fig:rates_and_Svac} we also plot the enhanced degree of discrimination for 3MCP inside a medium with and without local field corrections. In the next section we derive the necessary formulas to consider such a surrounding medium and discuss the resulting effect on the discrimination.
\section{Enhanced discrimination in magnetodielectric medium}
\label{sec:med}
%\begin{figure}[t]
%\includegraphics[width = 0.9 \linewidth]{non-chiral_bounds_in_nr.pdf} 
%\includegraphics[width = 0.9 \linewidth]{nrmax_non-chiral-med.pdf}
%\caption{\textbf{upper:} Upper and lower bound of $S$ for 3MCP in a dielectric medium. \textbf{lower:} Refractive index of a dielectric medium that leads to maximum discrimination as a function of the molecule's electric transition dipole divided by its magnetic one. We assumed that both donor and acceptor are the same molecules. The solid line marks the ratio for 3MCP.}
%\label{fig:Smax}
%\end{figure}
While in free space the degree of discrimination is completely determined by the molecules involved and their separation distance, we may modulate the effect by introducing a medium that surrounds the molecules. This can be either a liquid or gas that is well described by its macroscopic properties. Let us consider a magnetodielectric medium with relative permittivity and permeability $\varepsilon(\omega),\mu(\omega) \neq 1$ that are in general complex-valued, see Fig.~\ref{fig:schemes}. We can easily include the impact of such a medium onto the excitation propagation via the appropriate Green's tensor \eqref{app:Gmed} and the rate formula \eqref{dual:rate}. The Green's tensor that fulfils the Helmholtz equation inside such a medium is given by 
\begin{align}
\G ( \bm r_2, \bm r_1, \omega ) 
&= 
- \frac{\mu  }{3 k^2 } \bm \delta(\bm r)
-
\frac{\mu \e^{\mi k r }}{4 \pi k^2 r^3}
%\nn
%&
%\quad 
%\times
\Big\{ 
\left[  1 - \mi k r - 
(k r)^2 \right] \ten{I} 
\nn
& \qquad
- 
\left[
3 - 3 \mi k r - (k r)^2 
 \right] \bm e_r \otimes \bm e_r
\Big\}
,
\end{align}
with $\bm r = \bm r_2 - \bm r_1$, $k = \sqrt{\varepsilon \mu } \, \omega /c$ and $\bm e_r = \bm r/ r$. Using the definition of $S$, and evaluating the retarded and nonretarded limits, we find 
\begin{align}
S_{r \rightarrow 0} &= \frac{4  R_\mathrm{D} R_\mathrm{A} \left( \re[n_r]^2 - \im[n_r]^2 \right) }{ c^2 (|\bm d_\mathrm{e}^\mathrm{A} |^2|\bm d_\mathrm{e}^\mathrm{D} |^2 + |n_r|^4|\bm d_\mathrm{m}^\mathrm{A} |^2|\bm d_\mathrm{m}^\mathrm{D} |^2)} 
,
\\
S_{r \rightarrow \infty} &=
\frac{8  R_\mathrm{D} R_\mathrm{A} \re[n_r]^2}{c^2 \left(  |\bm d^\mathrm{A}_\mathrm{e} |^2 + |n_r|^2 |\bm d^\mathrm{A}_\mathrm{m}|^2 \right)\left(  |\bm d^\mathrm{D}_\mathrm{e} |^2 + |n_r|^2 |\bm d^\mathrm{D}_\mathrm{m}|^2 \right) } 
,
\end{align}
where $n_r = \sqrt{\varepsilon \mu }$ is the medium's complex refractive index, and in contrast to the free space case we did not neglect higher orders in $\bm d_\mathrm{m}$ to account for cases with $|n_r| \gg 1$. The free space discrimination \eqref{Sfs} can be recovered for $n_r = 1$.

\begin{figure}[t]
\includegraphics[width = 0.9 \linewidth]{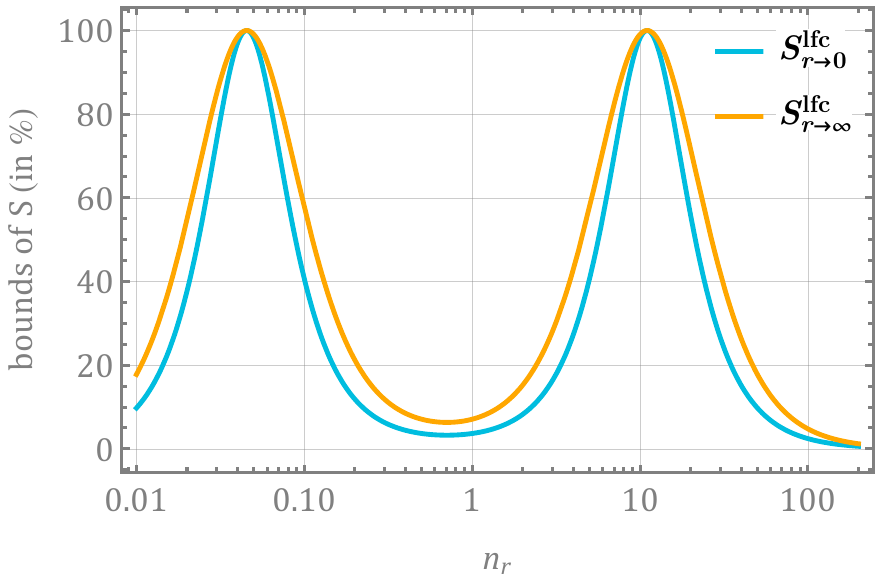} 
\includegraphics[width = 0.9 \linewidth]{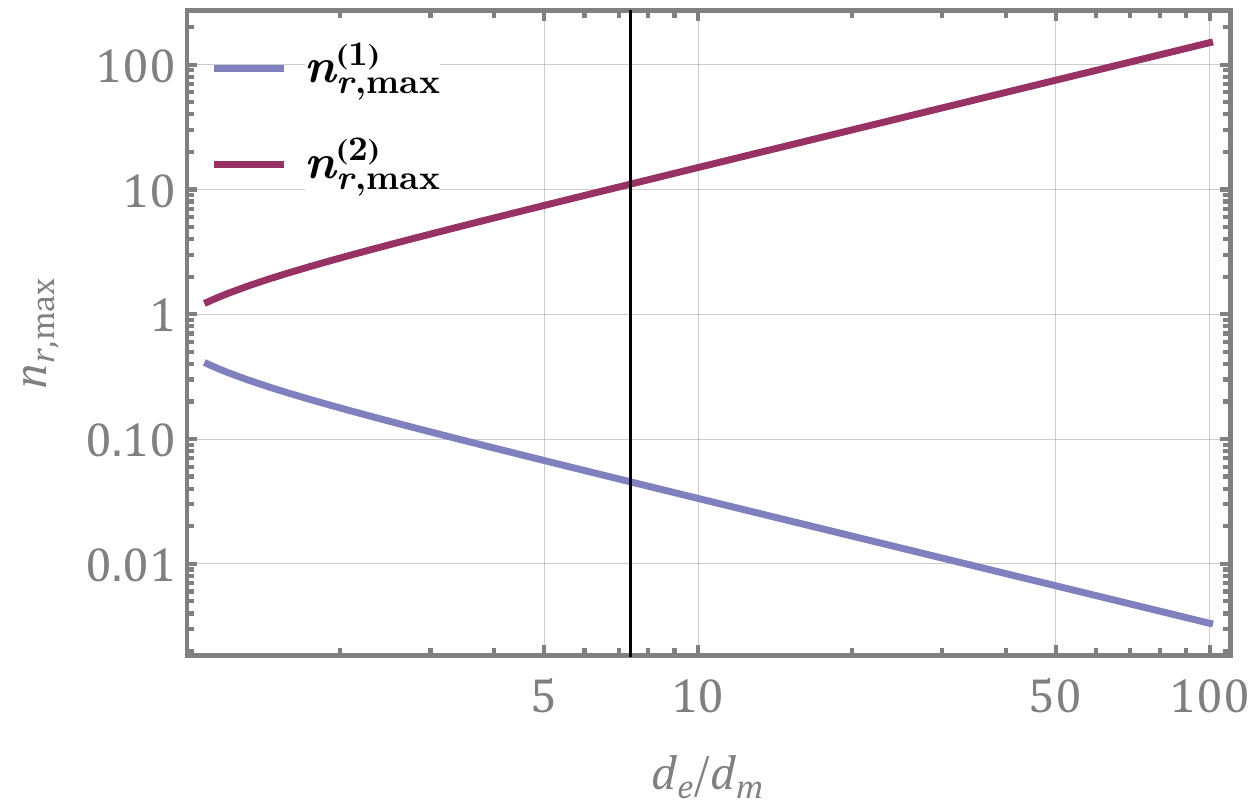}
\caption{\textbf{upper:} Upper and lower bound of $S$ for 3MCP in a dielectric medium. \textbf{lower:} Refractive index of a dielectric medium that leads to maximum discrimination as a function of the molecule's electric transition dipole divided by its magnetic one. We assumed that both donor and acceptor are the same molecules. The solid line marks the ratio for 3MCP.}
\label{fig:Smax}
\end{figure}
\begin{figure}[t]
\includegraphics[width = 0.9 \linewidth]{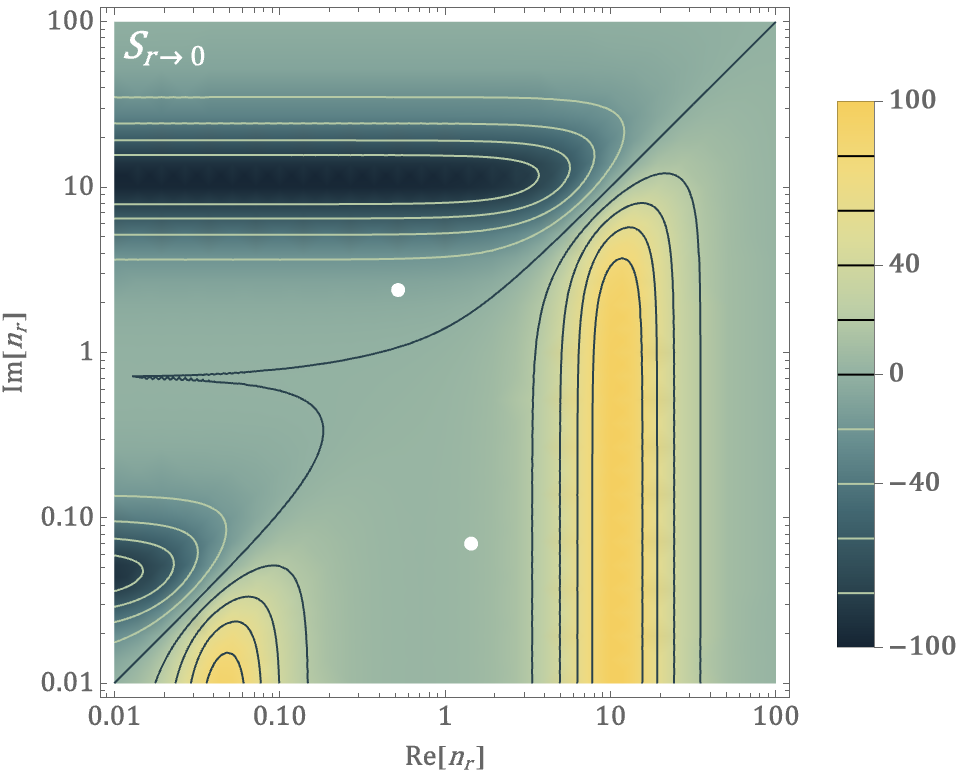}
\includegraphics[width = 0.9 \linewidth]{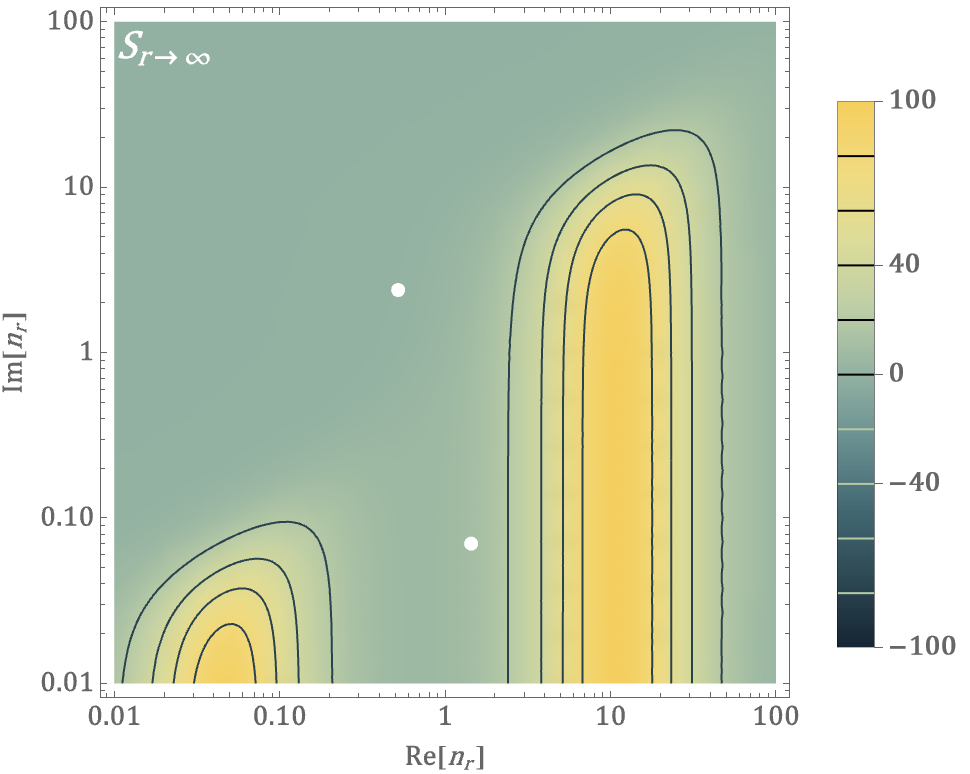}
\caption{Nonretarded (upper) and retarded (lower) degree of discrimination as a function of real and imaginary part of the refractive index (with $\mu =1$) for the example of 3MCP. We marked simple media with complex $n_r$: Methane at $n_r = 1.44 +  0.04 \mi$, mercury at $n_r = 0.52 + 2.39\mi $.}
\label{fig:boundsMed}
\end{figure}

However, it is known that this description is overly simplified. When embedding the microscopic system into the macroscopically described medium local field effects around donor and acceptor as well as screening effects must be taken into account. Without these, the model would assume emission and absorption inside of the macroscopic medium, which is incorrect as the interactions with the field appear inside of the microscopically described molecules that are not permeated by a continuous medium.
This so-called local field correction is in fact a necessity to correct the naive approach. There are a variety of different local field models \citep{Duan2005,Fiedler2017}. Here we employ the Onsager real cavity model \citep{Onsager1936}. It mainly introduces infinitesimal vacuum spheres around emitter and absorber. In the case of a homogeneous magnetodielectric medium the correction leads to additional factors to the dual Green's tensors. They differ depending on the nature of the interaction at each point (electric or magnetic). The local field corrected dual Green's tensor reads
\begin{align}
\G^\lfc_{\lambda \lambda'} 
&= c_\lambda \G_{\lambda \lambda'} c_{\lambda'} 
,
\end{align}
with
\begin{align} 
c_\mathrm{e} 
& = \frac{3 \varepsilon}{1 + 2 \varepsilon},  
\quad
\text{and} \quad 
c_\mathrm{m} 
= \frac{3 }{1 + 2 \mu}
.
\end{align}
Using $\G^\lfc_{\lambda \lambda'}$ in the rate formula  \eqref{dual:rate} we find the rate in such a medium including local field effects. 
The discriminatory and non discriminatory rate contributions are then given by 
\begin{widetext}
\begin{align}
\Gamma_\mathrm{disc}
&= 
\frac{R_\A R_\D  \omega ^2 | \mu  | ^2 e^{-2 \im n_r k_0 r  } }
{18 \pi  c^4 k_0^2 r^6 \varepsilon_0^2 \hbar ^2 | n_r | ^4} 
\bigg\{
k_0^2 r^2 | c_\e| ^2 | c_\m| ^2 | n_r | ^4 
\left(k_0^2 r^2 | n_r | ^2+2 k_0 r  \im n_r +1\right)
\nn&	\qquad\qquad\qquad	\qquad\qquad\qquad
+ \re \left[ c_\e^{*2} c_\m^2 n_r ^2\right] \Big(k_0^2 r^2 | n_r | ^2 (2 k_0 r  \im n_r  +1)+k_0^4 r^4 | n_r | ^4+k_0^2 r^2 \im n_r ^2
\nn
&	\qquad\qquad\qquad	\qquad\qquad\qquad\qquad
\qquad\qquad\qquad	\qquad\qquad\qquad\qquad
\qquad\qquad\qquad	\qquad\qquad
+6 k_0 r \im n_r  +3\Big)
\bigg\}
,
\label{Gammadisc}
\end{align}
\begin{align}
\Gamma_\mathrm{nd}&=
\frac{| \mu| ^2 | \bm d_\e^\A| ^2 | \bm d_\e^\D| ^2  e^{-2 k_0 r (\im n_r) }
}{36 \pi   r^6 \varepsilon_0^2 \hbar ^2 | n_r| ^4}
\bigg\{
	| c_\e| ^4 
	\Big(
		| n_r| ^4 k_0^4 r^4 
		+| n_r| ^2 k_0^2 r^2 (2 \im n_r k_0 r +1) 
		+4 \im n_r^2 k_0^2 r^2 
		+6 \im n_r k_0 r +3
	\Big)  
\nn&	\qquad\qquad\qquad	\qquad\qquad\qquad\qquad
	+
	| c_\e| ^2 |n_r c_\m| ^2  
	\Big(
		\frac{ | \bm d_\m^\D| ^2}{| \bm d_\e^\D| ^2 }
		+\frac{| \bm d_\m^\A| ^2 }{| \bm d_\e^\A| ^2 }
	\Big)
	\Big(
		 2 | n_r| ^4 k_0^4 r^4  +| n_r| ^2  k_0^2 r^2 
		(2  \im n_r k_0 r+1)
	\Big)
\nn&	\qquad\qquad\qquad	\qquad\qquad\qquad\qquad
+
		| n_r c_\m| ^4 \frac{ | \bm d_\m^\D| ^2| \bm d_\m^\A| ^2 }{| \bm d_\e^\D| ^2| \bm d_\e^\A| ^2} 
	\Big(
		| n_r| ^2  k_0^2 r^2 
		(2  \im n_r k_0 r+1)+| n_r| ^4 k_0^4 r^4 
		+4 \im n_r^2 k_0^2 r^2
\nn&	\qquad\qquad\qquad	\qquad\qquad\qquad\qquad
\qquad\qquad\qquad	\qquad\qquad\qquad\qquad
\qquad\qquad\qquad	\qquad\qquad
+6\im n_r k_0 r +3
	\Big)
\bigg\}
,
\label{Gammand}
\end{align}
where $k_0 = \omega/c$, and $\Gamma_\mathrm{nd}$ reduces to the known electric dipole--dipole RET-rate for $|\bm d_m| = 0$ \citep{Juzeliunas1994,Andrews2020}.
\end{widetext}
By dividing Eq.~\eqref{Gammadisc} by Eq.~\eqref{Gammand} and performing the limit of small and large distances, respectively, we obtain the degree of discrimination in its retarded and nonretarded limit. For general magnetodielectric media including local field effects they are given by
\begin{align}
S^\lfc_{r \rightarrow 0}
&= \frac{4  R_\mathrm{D} R_\mathrm{A}}{D^\lfc_0} 
\re\left[ (c_\mathrm{e}^* c_\mathrm{m} n_r)^2 \right] 
,
\\
D_0^\lfc &= c^2 \big(|c_\mathrm{e}|^2 |\bm d_\mathrm{e}^\mathrm{A} |^2|\bm d_\mathrm{e}^\mathrm{D} |^2  
\nn
& \qquad\qquad\qquad
+|c_\mathrm{m}|^2 |n_r|^4|\bm d_\mathrm{m}^\mathrm{A} |^2|\bm d_\mathrm{m}^\mathrm{D} |^2
\big)
,
\label{D0}
\\
S^\lfc_{r \rightarrow \infty}
&=
\frac{8  R_\mathrm{D} R_\mathrm{A}}{D^\lfc_\infty}  
\re\left[ c_\mathrm{e}^* c_\mathrm{m} n_r \right]^2
,
\\
D_\infty^\lfc
&=
 c^2 \left( |c_\mathrm{e}|^2 |\bm d^\mathrm{A}_\mathrm{e} |^2 + |c_\mathrm{m}|^2|n_r|^2 |\bm d^\mathrm{A}_\mathrm{m}|^2 \right) 
 \nn
& \qquad\qquad 
 \times
 \left(  |c_\mathrm{e}|^2 |\bm d^\mathrm{D}_\mathrm{e} |^2 + |c_\mathrm{m}|^2|n_r|^2 |\bm d^\mathrm{D}_\mathrm{m}|^2 \right)  
,
 \label{Dinfty}
\end{align}
where for each electric and magnetic transition dipole moment there appears an electric correction factor $c_\mathrm{e}$ and magnetic correction $c_\mathrm{m}$, respectively. The uncorrected case can be recovered for $c_\mathrm{e} = c_\m =1$ and the free space case for $\varepsilon = \mu = n_r = 1$.
In Fig.~\ref{fig:SmedLFC} the degree of discrimination is shown with and without correction for water with real permittivity $\varepsilon(\omega) \approx 1.82$ and trivial permeability $\mu(\omega) = 1$ (i.e.~ $n_r \approx 1.35$). As demonstrated in Fig.~\ref{fig:rates_and_Svac} for our example, the impact of the local field correction is of similar magnitude as the impact of the medium itself compared to the free-space case. Without corrections, the medium's impact on the degree of discrimination would be overestimated here. 
However, even in water the degree of discrimination is enhanced by roughly 30\%. 
An appropriate medium may enhance the degree of discrimination in general up to $S =  \prod_\mathrm{X}  \im [\bm d_\m^\mathrm{X} \cdot \bm d_\e^\mathrm{X}]/ (|\bm d_\m^\mathrm{X}| |\bm d_\e^\mathrm{X}|) = \prod_\mathrm{X} \cos \theta_\mathrm{X} \le 100\%$, where $\mathrm{X} =$A,D. A degree of discrimination of 100\% corresponds then to vanishing excitation transfer to the opposite-handed acceptor ($\Gamma_\mathrm R = 0$ for a left-handed donor) independent of the separation distance. It can only be achieved for $\theta = 0$. This is approximately the case in the chosen example of 3MCP. If we limit ourselves to positive real refractive indices $n_r > 0$ with trivial permeabilities $\mu \approx 1$, the maximum enhancement can be achieved for 3MCP at $n_{r,\text{max}} \approx 11$, and alternatively but less relevant, for $n_{r,\text{max}} \approx 0.045$, see Fig.~\ref{fig:Smax}. It is interesting to note that the discrimination vanishes in the limit $n_r \rightarrow \infty$. 
We can predict the maximum real refractive indices at which $S= \cos^2\theta$ for any two chiral molecules of the same kind as a function of the ratio $|\bm d_\e|/ |\bm d_\m|$, as shown in Fig.~\ref{fig:Smax}. 
The smaller the ratio the closer both optimum refractive indices are to unity. 
In Fig.~\ref{fig:boundsMed} we present the nonretarded and retarded degree of discrimination $S_{r \rightarrow 0/\infty}$ as functions of a complex refractive index. In theory, one may achieve a complete inversion of the discriminatory effect in the nonretarded limit for $\im n_r \gg \re n_r$.
%In the retarded case a higher imaginary part only results in vanishing discrimination. However, in the nonretarded regime the discrimination can be inverted for $\im n_r > \re n_r$.
In the chosen example of 3MCP the same-handed rate vanishes, $\Gamma_\mathrm L = 0 \Leftrightarrow S_{r\rightarrow 0} = - 100\%$, for $\im n_r = 11$ and $\im n_r = 0.045$. However, the retarded limit does not experience such an inversion of the effect. Here the discrimination simply vanishes with larger $\im n_r$. 
Most conventional media may be found around $1 < \re n_r < 2$ and $0 < \im n_r \ll 1$.
At the transition frequency of 3MCP ($\omega_\D = 6.44 \times 10^{15}$ s$^{-1} = 4.3$ eV) some example media are given in Table~\ref{tab:media} \citep{rii}. If possible, their refractive indices were marked in Fig.~\ref{fig:boundsMed}. An interesting simple medium example is mercury where the discrimination is inverted in the nonretarded limit compared to the free space case.
Solutions with resonances at the desired frequency as well as fluids based on metamaterials may be engineered to cover a larger range of the presented parameter space \citep{Zhang2018}. The choice of an appropriate medium then depends highly on the transition frequency. 
\begin{table}
\begin{tabular}{  r||c|c|c  } 
\, medium\,&\,  refractive index \, &$\quad S_{r\rightarrow 0}\quad$&$\quad S_{r\rightarrow \infty}\quad$\,  \\ 
\hline 
water\,& $1.4 + 10^{-8} \mi$ & 5.04\% & 9.60\% \\  
biodiesel\,& $1.56 + 8 \times 10^{-6} \mi$ & 5.85\% & 11.06\%  \\  
ethanol\,& $ 1.39 + 3 \times 10^{-6} \mi$ & 5.01\% & 9.55\% \\  
methane\,&$1.44 + 0.07 \mi$ & 5.19\% & 9.87\%\\  
mercury\,& $0.52 + 2.39 \mi$ & -7.30\% & 0.96\%\\   
\end{tabular} 
\caption{table of different simple media, their respective refractive index and their degree of discrimination in the nonretarded and retarded limit \citep{rii}. }
\label{tab:media}
\end{table}
\section{Conclusion}
\label{sec:conclusion}
We derived the RET rate between chiral molecules in the framework of macroscopic QED, that includes retardation effects and is able to take the impact of different environments into account. 
Although the considered system consists of one donor and one acceptor molecule, the extension to $N$ molecules is straightforward. 
The rate differs between same- and opposite-handed enantiomers, making the rate discriminatory. 
The degree of discrimination is usually larger in the far-zone, also known as the retarded limit of large molecule separation distances.  However, this constitutes a trade-off between the degree of discrimination and the overall energy transfer rate, which decreases rapidly with increasing separation distance.  
We showed that putting the system inside a magnetodielectric medium can have significant effects on the discrimination. 
In one example we demonstrated the significant impact of local field and screening effects on the prediction of the degree of discrimination. Here, we chose the Onsager real cavity model to account for these effects. 
We offered analytical expressions for the rate involving chiral molecules in a magnetodielectric medium including local field effects. We applied our theory to the example of 3-methylcyclopentanone (3MCP). By putting the system inside water the degree of discrimination between two chiral 3MCP molecules is already enhanced by roughly 30\%. 

We studied a large complex parameter space for dielectric media and showed that appropriate media may even lead to perfect discrimination, i.e.~the energy transfer to the opposite-handed enantiomer is completely suppressed. The degree of discrimination shows in general two such maxima for real refractive indices. 
One of these optimal refractive indices is much larger than unity, the other much smaller, depending on the transition dipoles involved in the process. The smaller the ratio between electric and magnetic transition dipole moments the closer the optimal refractive indices are to unity. 
Media with large imaginary refractive index can even invert the effect in the near-zone or non-retarded limit of small separation distances, such that the opposite-handed enantiomer instead of the same-handed one is preferred by the process. 
The medium's macroscopic properties are evaluated at the molecule's transition frequency. Hence, the best choice for a medium depends on the molecules of interest. 
We offered some simple example media at the transition frequency of 3MCP. The transition frequency of the chosen example molecules is quite large and simple liquids show little optical response. However, for vibrational  transitions involving wavelengths in the IR regime, recent advancements in the field of metamaterials could soon offer media that explore a larger area of the presented parameter space. 

In this work we chose chiral molecules and an achiral environment. However, even macroscopic media can have chiral features. Although the environment only passively takes part in the energy transfer process, its chiral property might be able to actively discriminate enantiomers. Here, we employed the Onsager real cavity model to account for local field and screening effects. For some systems alternative local field models are more suited. Their influence on the discrimination will be studied in future work. 

\clearpage
\appendix
\section{Green's tensor}
\label{app:G}
The Fourier components of the electric field are given by Eq.~\eqref{Ewfield} with the shorthand notation
\begin{align}
\G_\e(\bm r, \bm r', \omega)
&= \mi \frac{\omega^2}{c^2} \sqrt{\frac{\hbar }{\pi \varepsilon_0} \im \varepsilon(\bm r', \omega)} \G(\bm r, \bm r', \omega)
\\
\G_\m(\bm r, \bm r', \omega)
&= 
\mi \frac{\omega}{c}
\sqrt{\frac{\hbar }{\pi \varepsilon_0} \frac{\im \mu(\bm r', \omega)}{|\mu(\bm r', \omega)|^2}}
\end{align}
Note that they are not related to the dual definition $\G_{\lambda\lambda'}$ for the Green's tensor. 
From this it follows that the Green's tensor $\G$ must fulfil the Helmholtz equation
\begin{multline}
\left[\nablaL_a \times \frac{1}{\mu(\bm r_a, \omega)} \nablaL_a \times - \frac{\omega^2}{c^2} \varepsilon(\bm r_a, \omega) \right] \G ( \bm r_a, \bm r_b, \omega ) 
\\
= \bm \delta(\bm r_a - \bm r_b) 
\end{multline} 
When considering homogeneous media it is sufficient to solve the scalar Helmholtz equation
\begin{align}
- \left[ \Delta_a + k^2 \right] g(\bm r_a , \bm r_b , \omega) = \delta(\bm r_a - \bm r_b)
\end{align}
with $k^2  = \epsilon \mu \omega/ c$ 
and the Green's tensor is then given by the Green's function $g$ as
\begin{align}
\G(\bm r_a, \bm r_b, \omega)
= 
\mu 
\left[ 
\ten{I}
+ \frac{1}{k^2} \nablaL_a \otimes \nablaL_a
\right]
g(\bm r_a, \bm r_b, \omega)
\end{align}
The scalar Green's function is then given by
\begin{align}
g(\bm r_a, \bm r_b, \omega)
= 
\frac{\e^{\mi k |\bm r_a - \bm r_b|}}{4 \pi |\bm r_a - \bm r_b| }
\end{align}
and the Green's tensor reads
\begin{multline}
\G (\bm r_a , \bm r_b, \omega)= 
- \frac{\mu }{3 k^2} \bm \delta(\bm r)
- \frac{\mu \e^{\mi k r}}{4 \pi k^2 r^3} 
\big\{ 
\left[ 
1 - \mi k r - k^2 r^2 
\right] \ten{I}
\\
- \left[ 
3 - 3 \mi k r - k^2 r^2 
\right]
\bm e_r \otimes \bm e_r 
\big\}
\label{app:Gmed}
\end{multline}
with $\bm r = \bm r_a- \bm r_b$ and $\bm e_r = \bm r/ r$.
In free space ($\epsilon = \mu =1$) we hence find
\begin{align}
\G^{(0)} (\bm r_a , \bm r_b, \omega)
&= 
\left[ 
\ten{I}
+ \frac{c^2}{\omega^2} \nablaL_a \otimes \nablaL_a
\right]
\frac{\e^{\mi \omega |\bm r_a - \bm r_b|/c}}{4 \pi |\bm r_a - \bm r_b| }
\\
&=
- \frac{1}{3 k_0^2} \bm \delta(\bm r)
- \frac{ \e^{\mi k_0 r}}{4 \pi k_0^2 r^3} 
\nn
& \qquad \qquad 
\times
\big\{ 
\left[ 
1 - \mi k_0 r - k_0^2 r^2 
\right] \ten{I}
\nn
& \qquad \qquad \quad
- \left[ 
3 - 3 \mi k_0 r - k_0^2 r^2 
\right]
\bm e_r \otimes \bm e_r 
\big\} 
\label{app:Gfs}
\end{align}
\section{Derivation of transition matrix elements}
\label{app:sec:Mfi}
The formula for the transition matrix element \eqref{Mfi} with the interaction Hamiltonian \eqref{Hint}, the initial and final state given by $\ket{i} = \ket{1}_\mathrm{D} \ket{0}_\mathrm{A} \ket{ \{ \bm 0 \} }_F$ and $\ket{f} = \ket{0}_\mathrm{D} \ket{1}_\mathrm{A} \ket{ \{ \bm 0 \} }_F $ and using the introduced dual notation, see Eq.~ \eqref{dualdef1} leads to the expression: 
\begin{widetext}
\begin{multline}
M = - \int \!\! \dif \omega' \!\! \int\!\! \dif^3 r' \sum_{ \{\lambda\}} \Big[
\braket{g| \hat{\bm d}_{\lambda_1}^\D |e}_\D  
\cdot 
\frac{ \braket{ \{ 0\} |  \hat{\bm  E}_{\lambda_1} (\bm r_\D )| \bm 1' }_\F     
 \braket{ \bm 1'| \hat{\bm  E}_{\lambda_2} (\bm r_\A ) | \{ 0\} }_\F
}{ \hbar \omega' +\hbar \omega_\A }
\cdot 
\braket{1|  \hat{\bm d}_{\lambda_2}^{\A} | 0}_\A 
\\
+ \braket{1| \hat{\bm d}_{\lambda_1}^\A |0}_\A
\cdot 
\frac{ 
\braket{ \{ 0\} |  \hat{\bm  E}_{\lambda_1} (\bm r_\A )| \bm 1' }_\F     
 \braket{ \bm 1'| \hat{\bm  E}_{\lambda_2} (\bm r_\D ) | \{ 0\} }_\F
}{ \hbar \omega' - \hbar \omega_\D}
\cdot 
\braket{g|  \hat{\bm d}_{\lambda_2}^\D | e}_\D
\Big]
\label{app:Mfi}
\end{multline}
\end{widetext}
where $\hat{\bm E}_\e = \hat{\bm E}$ and $\hat{\bm E}_\m = c \hat{\bm B}$. It is trivial to write the matrix element in a non-dual formulation. However, the dual definitions offer a simplified notation.
By introducing the projection onto the frequency subspace: 
\begin{align}
\hat{P}(\omega) 
&= \sum_{n, \lambda} \int \dif^3 r \ket{\bm n _\lambda( \bm r, \omega) }\bra{\bm n _\lambda( \bm r, \omega) } 
\end{align}
each term in Eq.~\eqref{app:Mfi} can be written as: 
\begin{multline}
\int \!\! \dif \omega' \!\! \int\!\! \dif^3 r' \sum_{ \lambda'} 
\frac{ \braket{ \{ 0\} |  \hat{\bm  E}_{\lambda_1} (\bm r_\alpha)| \bm 1' } 
 \braket{ \bm 1'| \hat{\bm  E}_{\lambda_2} (\bm r_\beta) | \{ 0\} } 
}{ \hbar \omega' +\hbar \omega }   
\\
=
\int \!\! \dif \omega' \!\! \int\!\! \dif^3 r' 
\sum_{ \lambda',n} 
\frac{ 
\braket{  \{ 0\} | 
\hat{\bm  E}_{\lambda_1} (\bm r_\alpha)
\ket{\bm n'} \bra{\bm n'}
\hat{\bm  E}_{\lambda_2} (\bm r_\beta)  
 | \{ 0\}   } 
}{ \hbar \omega' +\hbar \omega } 
\\
=
\int \!\! \dif \omega'  
\frac{ \braket{ \{ 0\} |  \hat{\bm  E}_{\lambda_1} (\bm r_\alpha)\hat{P}(\omega') \hat{\bm  E}_{\lambda_2} (\bm r_\beta) | \{ 0\} } 
}{ \hbar \omega' +\hbar \omega } 
\\
=
\int \!\! \dif \omega_1 \int \!\! \dif \omega_2 
\frac{ \Braket{   \hat{\bm  E}_{\lambda_1} (\bm r_\alpha, \omega_1) \hat{\bm  E}^\dagger_{\lambda_2} (\bm r_\beta, \omega_2) }_\mathrm{vac}
}{ \hbar \omega_2 +\hbar \omega } 
\end{multline}
where $\braket{\, \cdot \, }_\mathrm{vac}$ is the field vacuum's expectation value.
The correlation functions $ \braket{   \hat{\bm  E}_{\lambda_1} (\bm r_\alpha, \omega_1) \hat{\bm  E}^\dagger_{\lambda_2} (\bm r_\beta, \omega_2) } _\mathrm{vac}$ can be evaluated by considering the fields expansion in terms of the Green's tensor \eqref{Efield} -- \eqref{Bwfield}. For the magnetic-electric correlation function \eqref{BEcorr} we find
\begin{multline}
  \braket{\hat{\vec{E}}_m (\bm r_a, \omega ) \otimes \hat{\vec{E}}^\dagger_e (\bm r_b, \omega' ) }
= c \braket{ \hat{\vec{B}} (\bm r_a, \omega ) \otimes \hat{\vec{E}}^\dagger (\bm r_b, \omega' )  } 
\\
= \sum_{\lambda, \lambda'} \iint\!\! \dif^3 r' \dif^3 r  \frac{c}{\mi \omega} 
\big\langle
\nablaL_a \times \G_\lambda ( \bm r_a, \bm r, \omega ) 
\cdot \hat{\vec{f}}_\lambda (\bm r, \omega)
\\
 \qquad\qquad \qquad\qquad 
\otimes \vec{\hat{f}}_{\lambda'}^\dagger (\bm r', \omega') 
\cdot \G _{\lambda'}^*(\bm r' , \bm r_b,\omega')  
\big\rangle
\\
=
\frac{\omega  c }{\mi } \frac{\hbar \mu_0 }{\pi} \delta( \omega - \omega') \nablaL_a \times \im \G(\bm r_a , \bm r_b, \omega)  
\end{multline}
the remaining correlation functions, see Eqs.~\eqref{EEcorr}-- \eqref{BBcorr} follow analogously.
\section{Pole integration}
\label{app:pole}
\begin{figure} 
\includegraphics[width =  \linewidth]{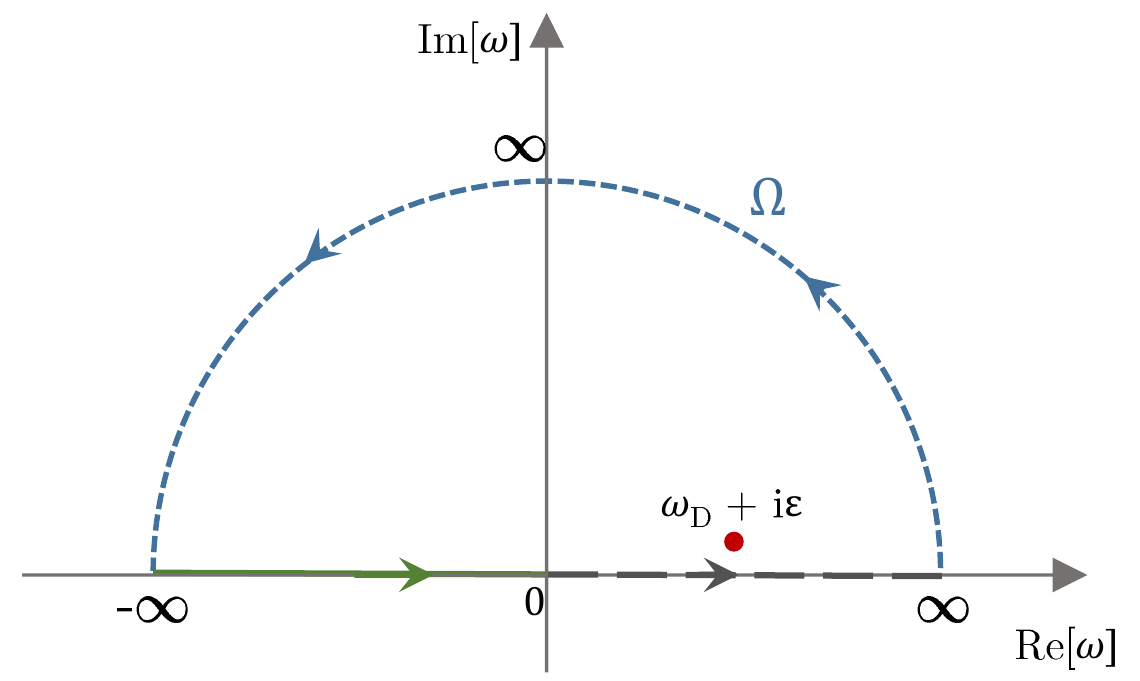} 
\caption{Complex contour for the pole integration. The closed contour consists of the sum of the green contour (from $- \infty$ to $0$), the gray contour (from $0$ to $\infty$) and blue dashed one ($\Omega$). It encloses a pole at $\omega = \omega_\D + \mi \epsilon$.}
\label{fig:cont}
\end{figure}   
Here we show the relation Eq.~\eqref{poleInt}.
With the Schwarz-reflection principle, $\G^*(\omega) = \G(- \omega^*)$ we find
\begin{multline}
\int_0 ^\infty \dif \omega \frac{f(\omega)\im G(\omega)}{ \omega - (\omega_\D + \mi \epsilon) } 
=
\frac{1}{2\mi}
\int_0^\infty 
\dif \omega f(\omega)  \frac{\G  (\omega)}{  \omega  - ( \omega_\D + \mi \epsilon)} 
\\
- 
\frac{1}{2\mi}
\int_0^\infty 
\dif \omega f(\omega)  \frac{\G  (-\omega)}{  \omega  - ( \omega_\D + \mi \epsilon)}   
\\
=
\frac{1}{2\mi}
\left[\oint -  \int_\Omega - \int_{-\infty}^0 \right]
\dif \omega f(\omega)  \frac{\G  (\omega)}{  \omega  - ( \omega_\D + \mi \epsilon)}   
\\
+
\frac{1}{2\mi}
\int^0_{-\infty}
\dif \omega f(-\omega)  \frac{\G  (\omega)}{  \omega  + ( \omega_\D + \mi \epsilon)}  
\end{multline}
where we expressed the integration as sum over different integration paths in the complex plane, see Fig.~\ref{fig:cont}. The closed path can be evaluated via the residuum theorem and the integration over path $\Omega$ vanishes. Similarly the second integral can be substituted by $\oint - \int_\Omega - \int_0^\infty$ the closed contour in this case does not enclose a pole and vanishes as well: 
\begin{multline}
\frac{1}{2\mi}
\left[\oint -  \int_\Omega - \int_{-\infty}^0 \right]
\dif \omega f(\omega)  \frac{\G  (\omega)}{  \omega  - ( \omega_\D + \mi \epsilon)}   
\\
+
\frac{1}{2\mi}
\int^0_{-\infty}
\dif \omega f(-\omega)  \frac{\G  (\omega)}{  \omega  + ( \omega_\D + \mi \epsilon)}  
\\
= 
\frac{1}{2\mi}
\left[2 \pi \mi  f(\omega_\D + \mi \epsilon) \G  (\omega_\D + \mi \epsilon)  \right]
\\
-\frac{1}{2\mi} \int_{-\infty}^0 
\dif \omega f(\omega)  \frac{\G  (\omega)}{  \omega  - ( \omega_\D + \mi \epsilon)}   
\\
-
\frac{1}{2\mi}
\int_0^{\infty}
\dif \omega f(-\omega)  \frac{\G  (\omega)}{  \omega  + ( \omega_\D + \mi \epsilon)}  
\\
%= 
%\frac{1}{2\mi}
%\left[2 \pi \mi  f(\omega_\D + \mi \epsilon) \G  (\omega_\D + \mi \epsilon)  \right]
%+
%\frac{1}{2\mi} \int^{\infty}_0 
%\dif \omega f(-\omega)  \frac{\G  (-\omega)}{  \omega  + ( \omega_\D + \mi \epsilon)}   
%\\
%-
%\frac{1}{2\mi}
%\int_0^{\infty}
%\dif \omega f(-\omega)  \frac{\G  (\omega)}{  \omega  + ( \omega_\D + \mi \epsilon)}  
%\\
=  
\left[ \pi   f(\omega_\D + \mi \epsilon) \G  (\omega_\D + \mi \epsilon)  \right]
\\
-
\frac{1}{2\mi} \int^{\infty}_0 
\dif \omega f(-\omega)  \frac{\im \G  ( \omega)}{  \omega  + ( \omega_\D + \mi \epsilon)}    
\end{multline}
This result is finite for $\epsilon \rightarrow 0$, we may hence form the limit. Plugging this result back into our original expression reveals 
\begin{align}
\int_0^\infty \dif \omega 
&  \bigg( \frac{f(\omega)  }{  \omega -  \omega_\D}  + \frac{f(- \omega) }{  \omega  + \omega_\A}  \bigg) \im \G   (\omega)  = \pi f(\omega_\D) \G(\omega_\D) 
\end{align}
for $\omega_\A = \omega_\D$.
\end{document}